\documentstyle[pra,aps,epsf,rotating]{revtex}

\begin{document}

\title{An insulating grid spacer for large-area MICROMEGAS chambers}

\author{
D.~Bernard, H.~Delagrange, D.~G.~d'Enterria, M.~Le Guay,
G.~Mart\'{\i}nez\thanks{Corresponding author: martinez@in2p3.fr },
M.J.~Mora, P.~Pichot, D.~Roy and Y.~Schutz}

\address{
SUBATECH (Ecole des Mines de Nantes, IN2P3/CNRS, Universit\'e de
Nantes), \\ BP20722, 44307 Nantes Cedex 3, France}

\author{
A. Gandi and R. de Oliveira}
\address{CERN, CH - 1211 Geneva 23, Switzerland}

\maketitle

\date{\today}

\begin{center}{Abstract}\end{center}
{
\it We present an original design for large area gaseous detectors based on the MICROMEGAS technology.
This technology  incorporates an insulating grid, sandwiched between the micro-mesh and the anode-pad plane,
which provides an uniform 200 $\mu$m amplification gap.
The uniformity of the amplification gap thickness has been verified.
The gain performances of the detector are presented and compared to the values obtained with detectors using cylindrical micro spacers.
The new design presents several technical and financial advantages.
}

~\\ ~\\

Particle detectors based on the MICROMEGAS technology
\cite{Giom96,Giom98,Char98,Barc99,Cuss98,Micr99,Micr00,Aphe01}
consist of a parallel-plate gas-chamber split into two asymmetric
stages by an intermediate electrode made out of a micro-mesh.
By applying appropriate voltages between the three electrodes,
a very high electric field ($\sim 30$ kV/cm) in the thinnest stage (the
``amplification gap'') and a low electric field (3 kV/cm) in the
thicker stage (the ``drift'' or ``conversion gap'')
can be simultaneously established.
When a charged particle traverses the conversion gap, it generates primary
electrons which are subsequently multiplied in the amplification gap.
The amplified electron cloud is collected on the
anode generating a fast electric signal.
The associated ion cloud is quickly collected on the micro-mesh,
whereas only a small fraction of the ion cloud enters into the conversion region.
Based on this technology, we had
earlier developed \cite{Aphe01} large area particle detectors.
In the current detector, the main modifications with respect our previous developments \cite{Aphe01} are:

\begin{itemize}
\item The total area of the anode plane has been enlarged to 400$\times$400 mm$^2$,
providing an active area of 387$\times$ 387 mm$^2$. The anode
electrode consists of a 1.0 mm thick printed circuit board. Its
inner surface is segmented in rectangular 12$\times$12 mm$^2$
gilded copper pads. The inter-pad width is 100 $\mu$m and the total
number of pads is 1024.


\item The electro-formed micro-mesh\footnote{BMC Industries, 278 East 7th Street, St. Paul, MN 55101, USA.}
consists of a 7 $\mu$m thick grid of 508$\times$508 mm$^2$ made of
pure Ni. The 106 $\mu$m squared holes grid are outlined by a
20 $\mu$m thick border of Ni in steps of 126 $\mu$m, i.e. 200 LPI
(lines per inch). The optical transparency reaches 70\%. The
micro-mesh is stretched on the Plexiglas frame whose height
defines the 6 mm thick (compared to the 3 mm thick in the earlier design)
conversion gap between the micro-mesh and the cathode plane.
The amplification gap between the micro-mesh and the anode plane has been doubled to  200 $\mu$m.


\end{itemize}

In this letter, two different designs to keep an uniform amplification gap of 200
$\mu$m
have been compared:
\begin{itemize}
\item {\bf Old design.}
The amplification gap is defined by cylindrical micro-spacers of 200 $\mu$m high and 250 $\mu$m
in diameter, glued on to the anode-pads with a pitch of 2 mm in both directions.

\item {\bf New design.}
The micro-spacers are replaced by an insulating grid sandwiched between
the micro-mesh and the anode plane (see Fig.1).
\end{itemize}

The new design presents several decisive technical but also financial
advantages when compared to the cylindrical micro-spacer design.
Indeed, since the grid is independent of the printed-circuit board,
boards and grids can be interchanged in case of malfunction of one of the
two elements.
Moreover, the printed-circuit board without cylindrical micro-spacers
is more robust.
This particular feature facilitates the production process, in particular the polishing
of the anode-pad surface.
In addition, the connectors can be soldered more easily to the back of the board
by standard industrial techniques, thus reducing the cost of the
detectors.
The choice of the grid material is dictated by the
insulation properties of the material and by its mechanical
properties:
It must be resistant to heat and rigid enough to be
manipulated.
A 200 $\mu$m thick FR4 grid of the same size as the printed board
(394$\times$394 mm$^2$) was chosen.
The grid consists of 1024 squared cells of 11.8 mm width
and 300 $\mu$m pitch (see Fig.2).
On the edge of the grid, 28 2.4 mm diameter holes provides wide opening
for the gas to flow inside the detector.

The critical parameter in the new design is the uniformity of
the amplification-gap across the grid cells.
We have therefore measured the amplification-gap thickness
at different positions across the  grid cell and
for various voltages applied to the micro-mesh.
For that purpose, we used a magnifying camera whose position in the
three spatial directions can be accurately measured.
With this technique the thickness can be determined with a 5 $\mu$m accuracy.
For non-zero voltages the mesh flattens against the grid
near its edge of the grid. For usual operating voltages from -440 V
to -600 V (establishing electric fields from 22 kV/cm to 30 kV/cm), the average
amplification gap thickness is $(203\pm 5)\mu$m and the difference
in the gap thickness between the edge and the center of a grid hole is
less than 10 $\mu$m at -440 V and less than 7 $\mu$m at -600V (see Fig.3).
Such differences do not strongly affect the gain uniformity along the grid cell.
In the worse case a gain variation of no more than 40\%
might occur. This would induce a minor impact on the detection of minimum ionizing particles.
At voltages above -700 V (electric
field larger than 35 kV/cm) the mesh is deformed by the electric
force and the amplification gap thickness reaches a minimum value
of (190$\pm$5) $\mu$m at the center of the cell.

The gain of this newly designed detector  has been measured for different
gas mixtures and micro-mesh voltages.
A relative determination of the gain has been obtained by measuring the electric current induced in the micro-mesh
when the detector is irradiated by a $^{90}$Sr radioactive source.
The source was placed on a 50 $\mu$m
mylar foil replacing the anode of the detector.
We have compared the two designs, cylindrical micro-spacers\footnote{In this design,
the $^{90}$Sr radioactive source was placed on a 8 $\mu$m thick micro-mesh Ni
foil.}
and insulating grid.
To eliminate most of systematic
differences, the measurements have been performed simultaneously,
the two detectors being supplied with gas from the same source.
To estimate the absolute gain in the amplification gap, we have
assumed that the number of electrons collected by the cathode is
equal to the number of ions collected by the micro-mesh.
The resulting electric current gives a relative measure of the detector gain and
the following expression gives an estimation of the gain in the amplification gap:
\begin{equation}
G = \frac{I}{e<N>A}
\end{equation}
where $I$ is the measured electrical current, $e$ the electron
charge, $A$ the activity of the source expressed in Bq
(5.6$\times 10^5$ Bq in the present measurement) and $<N>$ the
average number of primary electrons created in the 6 mm conversion
gap. This later number obviously depends on the gas mixture, being
$<N>=27$ for Ne+CO$_{2}$(5\%),
$<N>=29$ for Ne+CO$_{2}$(10\%) and
$<N>=56$ for Ar+CO$_{2}$(10\%).
The sensibility of the apparatus to the micro-mesh current was 1nA. The results (see Fig.4) indicate that the two designs behave similarly,
although for a given voltage systematically higher gains are reached.
with the micro-spacer design: about 10 V more are needed for the
grid design to reach the same gain as with the micro-spacers.
This could be related with a systematic difference of the amplification gap thickness between both detector designs.
As a matter of fact, a  variation on the micro-mesh voltage of 10 V represents a gain variation of around 40\%.
This gain difference between both designs would correspond
to a gap-thickness systematic difference of only 10 $\mu$m.
In addition, the electron source was placed on a mylar
foil 50 $\mu$m thick for the new design and on a Ni micro-mesh 8 $\mu$m thick for the old design.

%
%
%


In summary, we have presented a new design for a gas chamber based
on Micromegas technology, where the 200 $\mu$m uniform
amplification gap is provided by an insulating grid sandwiched by
the micro-mesh and the cathode-pad board.
The micro-mesh is not strongly deformed for usual operating electric
fields.
The measured gain in the amplification gap
for different gas mixtures and micro-mesh voltages shows results
similar to those obtained with cylindrical micro-spacers. The grid
design presents several decisive technical as well as financial
advantages when compared to the cylindrical micro-spacers design,
thus allowing for the industrial production of inexpensive particle detectors.

This work has been supported by the IN2P3/CNRS, France and the ``Conseil R\'egional des
Pays de la Loire'', France.

\newpage

\begin{figure}
{\par\centering
\resizebox*{0.85\textwidth}{!}{\includegraphics{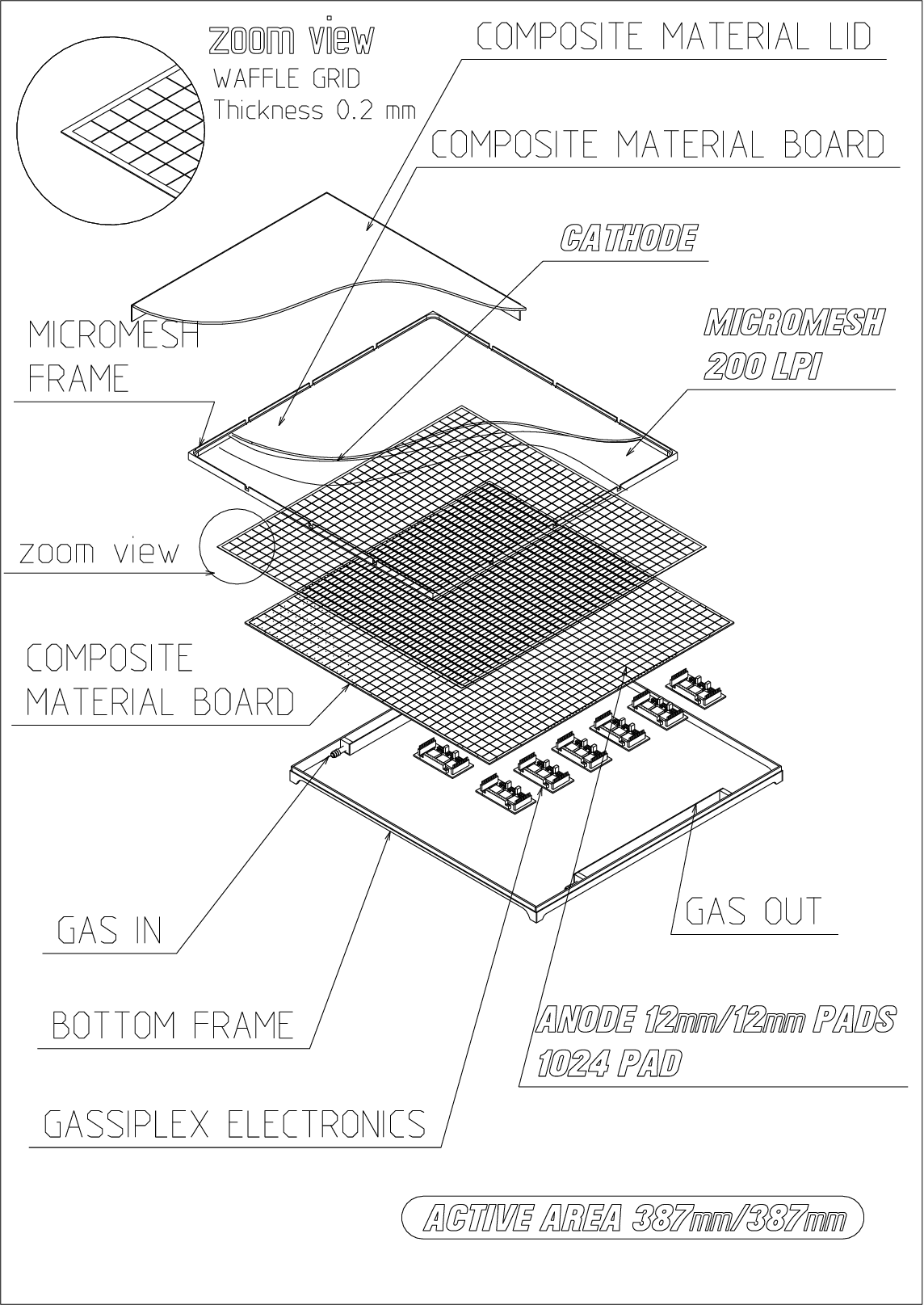}}
\par}
\vspace{0.8cm}
\caption{Sketch of the new design for a large area particle detector with
an insulating grid providing the amplification gap.}
\end{figure}

\newpage
\begin{figure}
{\par\centering
\resizebox*{0.82\textwidth}{!}{\includegraphics{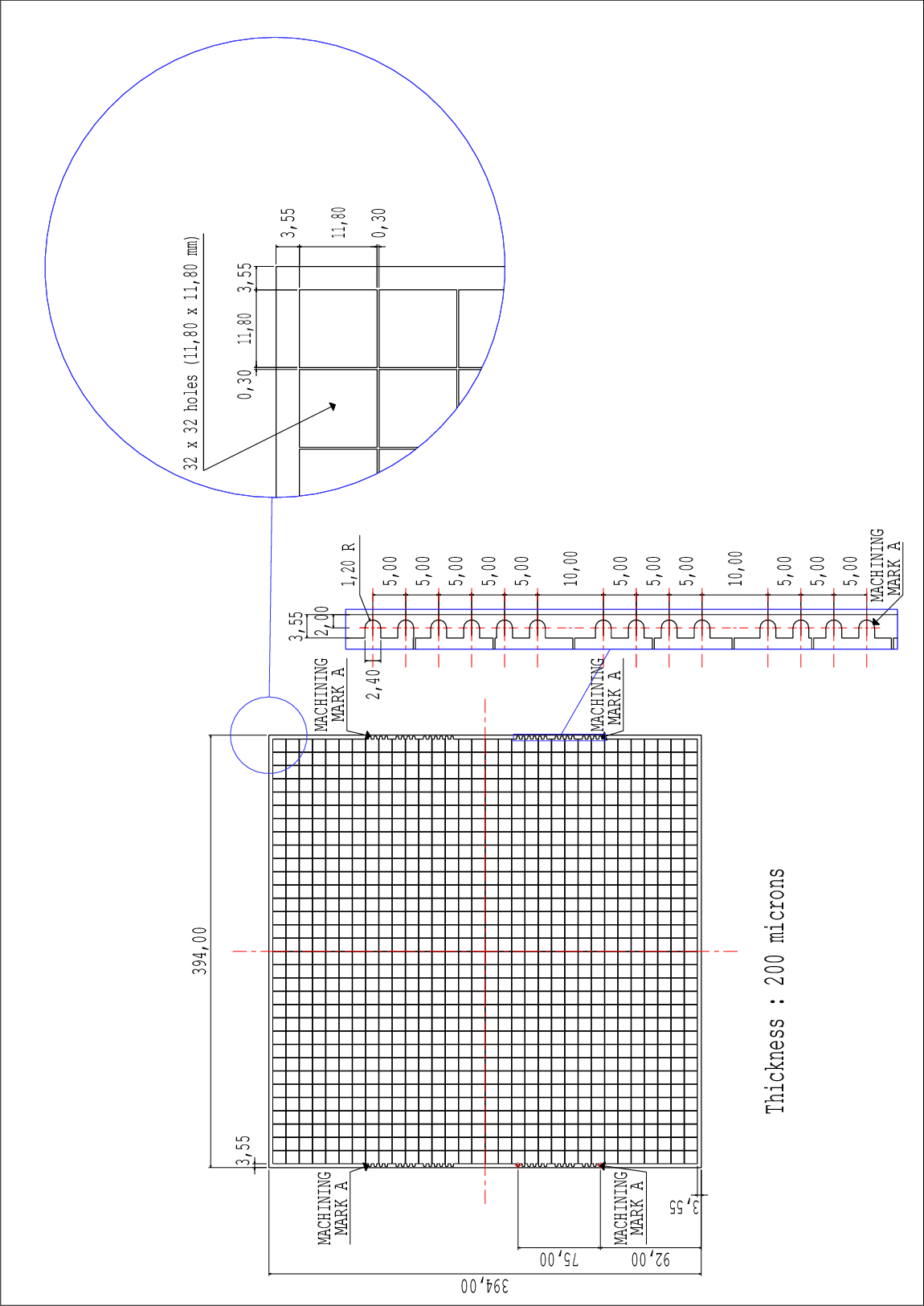}}
\par}
\vspace{0.8cm}
\caption{Technical drawing illustrating the machining applied on
a 200 $\mu$m insulating plate to obtain the grid. The
circular zoom details the rectangular holes which delimit active
areas. The rectangular zoom shows the holes which permit the gas to
flow freely inside the detector.}
\end{figure}


\newpage
\begin{figure}
{\par\centering
\resizebox*{0.95\textwidth}{!}{\includegraphics{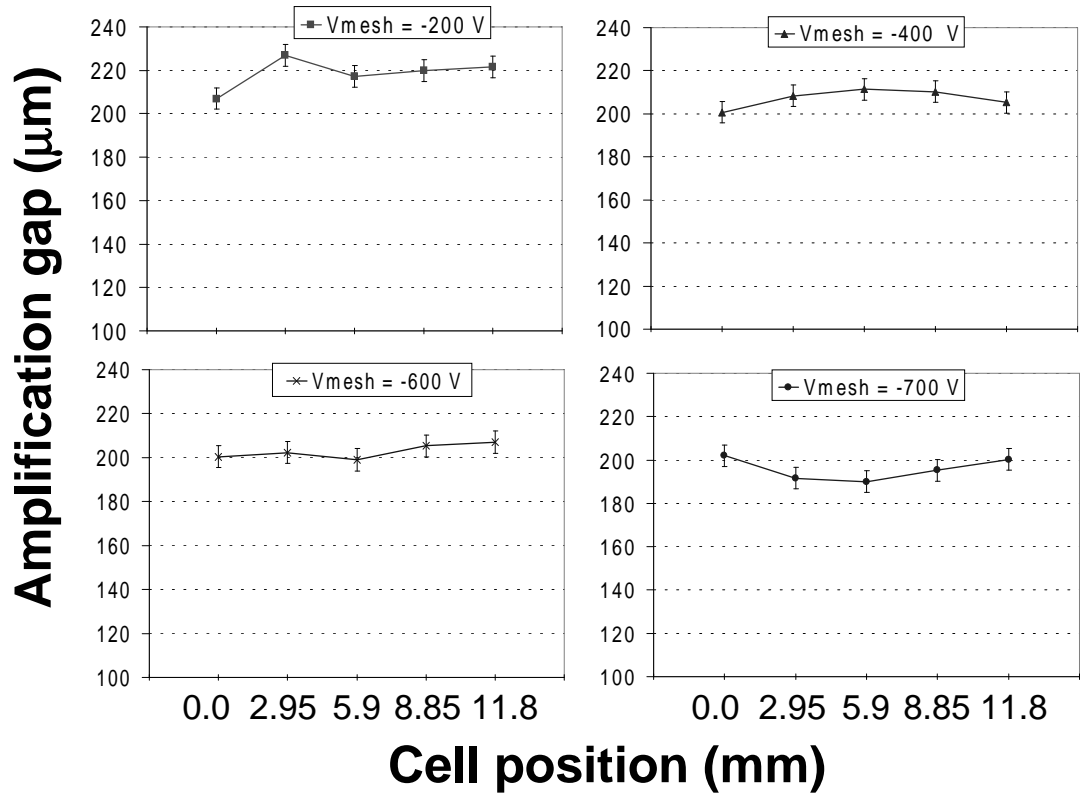}}
\par}
\caption{Amplification gap thickness on several points of the mesh across one hole
 of the grid as a function of voltage applied in the micro-mesh: -200 V, -400 V, -600 V and -700 V}
\end{figure}

\newpage
\begin{figure}
{\par\centering
\resizebox*{1.1\textwidth}{!}{\includegraphics{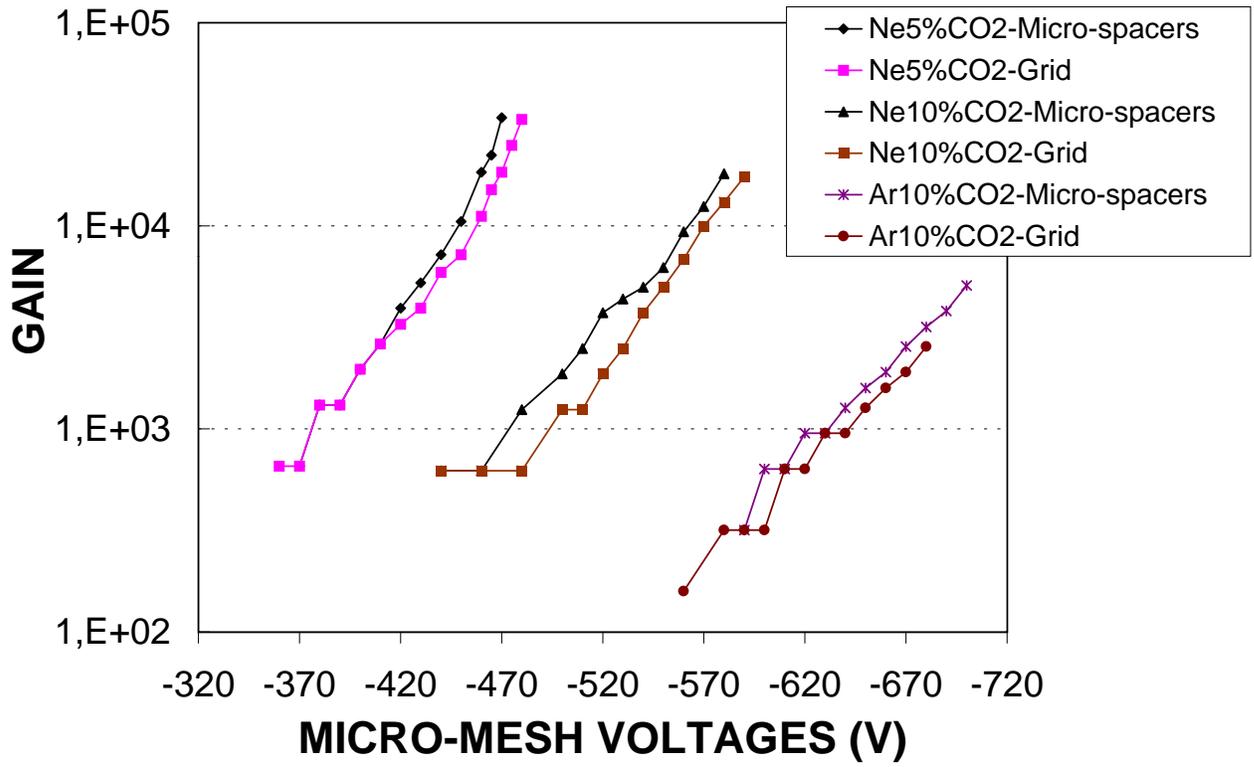}}
\par}
\caption{Measured gain as a function of the voltage applied to the micro-mesh
for two detector designs: one based on the micro-spacers design
and the insulating-grid design. Various gas mixtures have been
studied.}
\end{figure}

\end{document}